\documentclass{prop2015}
\usepackage[english]{babel}
\usepackage{hyperref}
\title{From black holes to flux throats} 
\subtitle{Polarization can resolve the singularity}
\author[D.\ Cohen-Maldonado]{Diego Cohen-Maldonado\inst{1}}
\author[J.\ Diaz]{Juan Diaz\inst{2}}
\author[T.\ Van Riet]{Thomas Van Riet\inst{2}}
\author[B.\ Vercnocke]{Bert Vercnocke\inst{1}\footnote{Corresponding author\quad E-mail:~\textsf{bert.vercnocke@uva.nl}}}
\address[1]{Institute of Physics, University of Amsterdam, Science Park,\\ 
	Postbus 94485, 1090 GL Amsterdam, The Netherlands }
\address[2]{ Instituut voor Theoretische Fysica, K.U. Leuven,\\
	Celestijnenlaan 200D B-3001 Leuven, Belgium}
	
\shortabstract
\begin{document}


\begin{abstract}
Supersymmetry-breaking is a key ingredient for string theory models to be phenomenologically viable. We review the strong analogy in the physics and the methods used for describing non-supersymmetric flux vacua and non-supersymmetric black holes in string theory. We also show how the polarized state could be the key to describing a well-behaved back-reaction of anti-branes in flux backgrounds, shedding a new light on a recent debate in the literature.
\end{abstract}


\maketitle

\section{Introduction}

In this note we look at two systems that are often studied to connect string theory, in  its capacity as a theory of quantum gravity, to the non-supersymmetric world.
One is the description of four-dimensional cosmological vacua in flux compactifications. The other is  ``zooming in'' on what are perhaps, from a theoretical perspective, the simplest gravitational solutions: black holes. 
 

In a maximally-symmetric four-dimensional  vacuum, physics is independent of the external space. A stationary black hole is similar to a flux compactification to a one-dimensional `vacuum': nothing depends on time \cite{Bena:2012ub}.  Note that the nine-dimensional `internal' space for a black hole is  non-compact:
\begin{equation}
\begin{array}{lcc}
\text{\bf Vacua:}& \overbrace{e^{2A(y)} g_{\mu\nu} dx^\mu dx^\nu}^{\text{4d vacuum}}
&
+\overbrace{g_{mn}(y) dy^m dy^n}^{\text{6d compact}}
\\
&\Updownarrow&\Updownarrow\\
\text{\bf BH's:}
&
\underbrace{-e^{2U(x)} dt^2}_{\text{1d time}}
&
+\underbrace{g_{mn}(x) dy^m dy^n}_{\text{3d external+ 6d compact}}
\end{array}
\label{eq:comparison}
\end{equation}

We have three goals with this paper. First, we want to review the similarities between both systems, on the level of their form (highly warped throats) and of the physics underlying their construction.  The mechanisms at play are non-trivial topology supporting flux and supersymmetry breaking by additional anti-branes. 

Second we stress the importance of general methods to gain insight in the behaviour of new supergravity solutions, without having to solve all equations of motion. Warped stationary throat regions carry a conserved mass that encodes the energy above the supersymmetric  minimum.  One can write the mass as a boundary term at spatial  infinity. By Stokes' theorem, it is a sum of integrals over a possible inner boundary $\partial \Sigma_{\rm int}$ and the bulk $\Sigma$:
\begin{equation}
M = \int_{\partial\Sigma} {\cal B} +\int_\Sigma {\cal C}\,,
\end{equation}
where the forms $\cal B, C$ depend on the fields of the theory.
This so-called \emph{Smarr relation} allows to confront the expected asymptotic behaviour of a solution to the near-anti-brane solution and shows whether near- and far solutions can be consistently glued. 

Third, we want to correct an error in our previous work \cite{Cohen-Maldonado:2015ssa} (version 1), where we used the Smarr relation to discuss the  back-reaction of the NS5  polarization channel of anti-D3 branes in warped throats that underlies certain vacua  constructions.\footnote{We thank Joe Polchinski and Andrea Puhm for pointing this out and for useful email correspondence.} After properly dealing with gauge field patches, we find that there is a unique radius for the polarized brane such that the $H_3$-flux singularity can be interpreted as the self-energy of the NS5  brane. However, the polarization radius does not seem to match the probe prediction of \cite{Kachru:2002gs}. We  point to  \cite{Cohen-Maldonado:2015ssa}, version 2, for more details.


\section{More than an analogy}\label{sec:analogy}

The similarity between flux vacua and black holes in string theory goes much further than the metric structure. We compare the presence of throats, the topological structure of supersymmetric solutions and the expectations for metastable solutions that break supersymmetry.

\subsection{Throats}

Flux vacua and black holes experience a region of infinite warping or redshift, where the warp factor of the world-volume  goes to zero. Flux compactifications of string theory with a region of high warping where $e^{2A}\to 0$ realize a  hierarchy of physical scales \cite{Giddings:2001yu}.  Also black holes curve spacetime to a maximum: at the horizon there is an infinite redshift and $e^{2U} \to 0$.  As we discuss now, the warping in string theory realizations leads to capped-off throats in which $e^{2U}$ and $e^{2A}$ can become very small  but remain non-zero.

\subsection{SUSY: topology and new phases}

\begin{figure}[ht!]
\centering
 \includegraphics[width=.39\columnwidth]{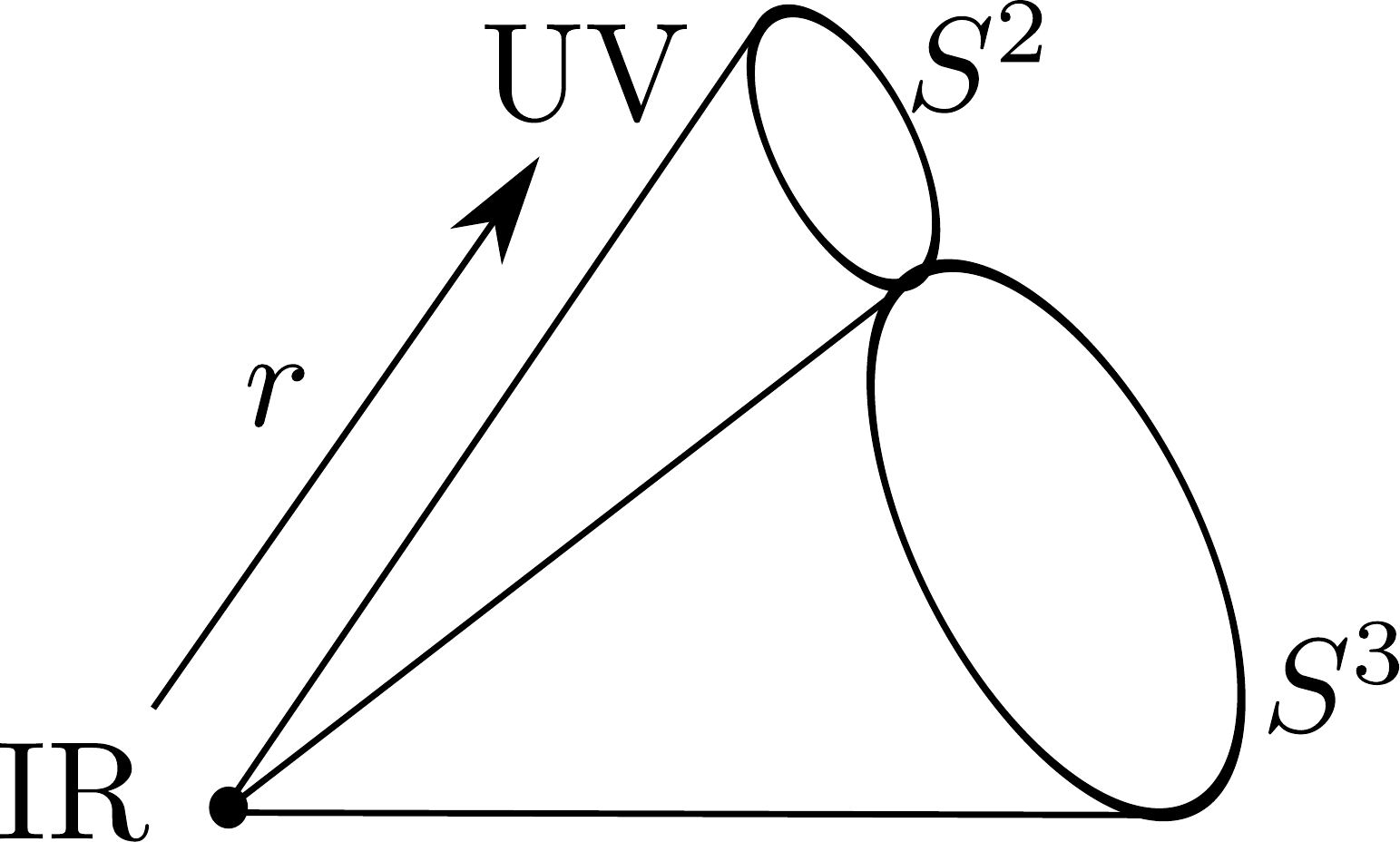}%
 \hspace{.18\columnwidth}
 \includegraphics[width=.34\columnwidth]{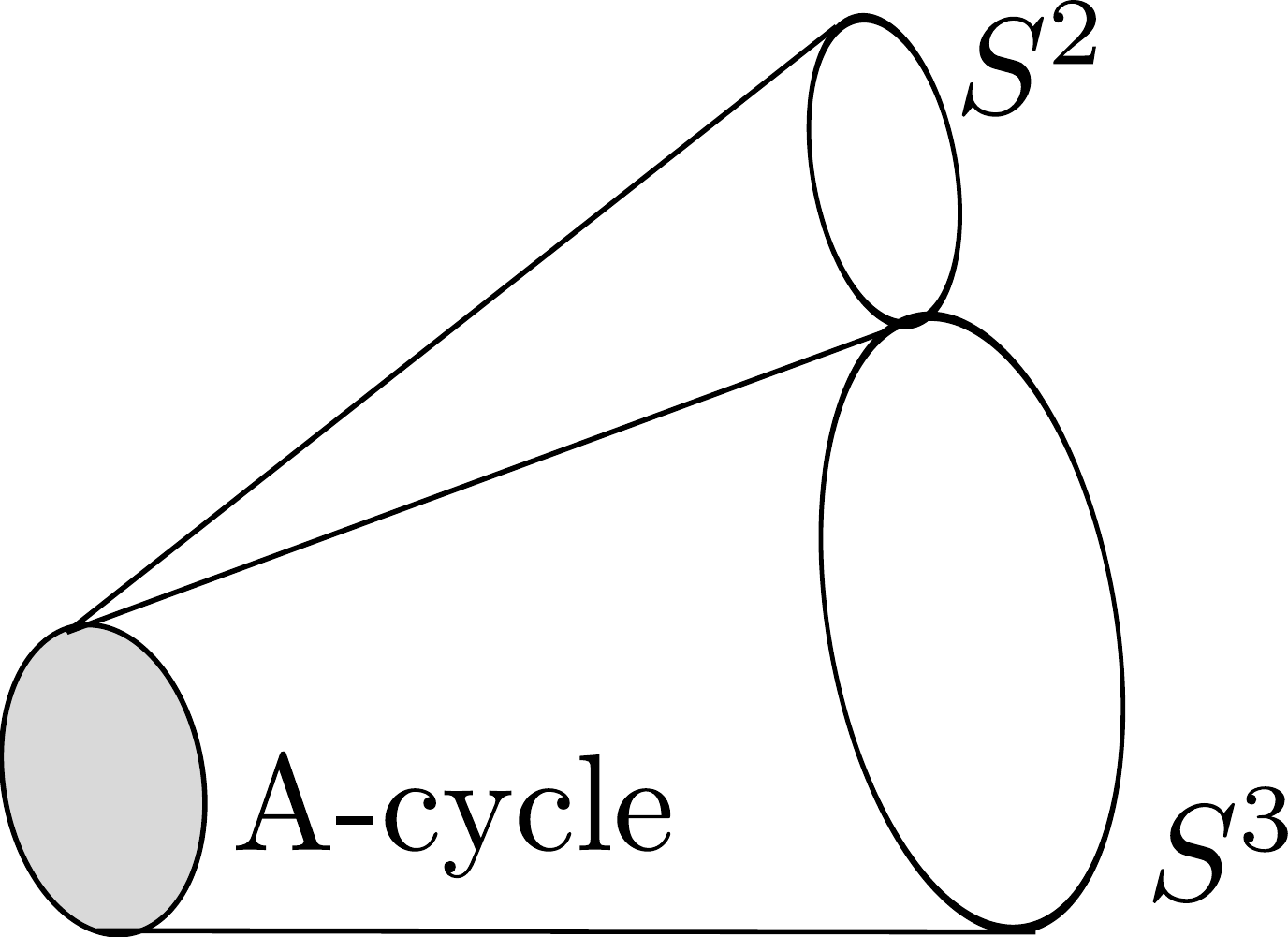}\\[8mm]
 \includegraphics[width=.35\columnwidth]{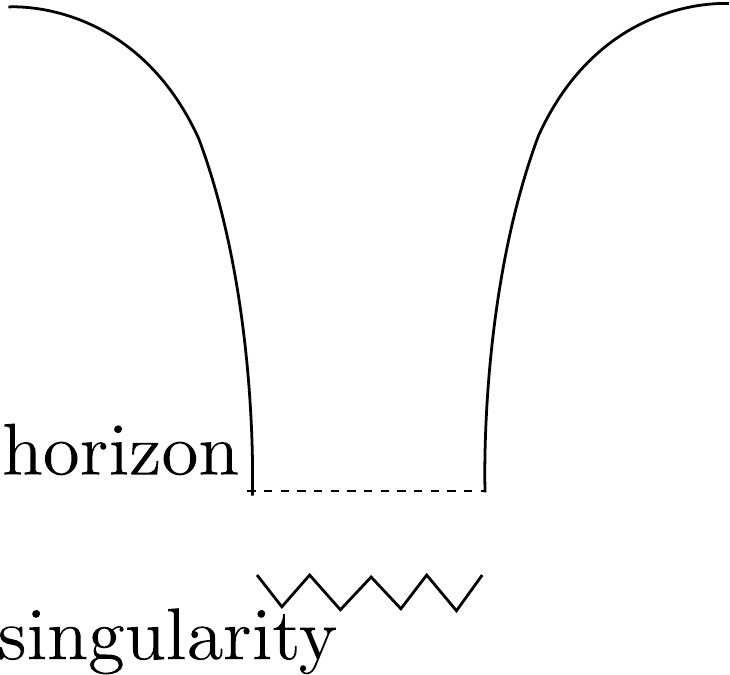}%
 \hspace{.18\columnwidth}
 \includegraphics[width=.3\columnwidth]{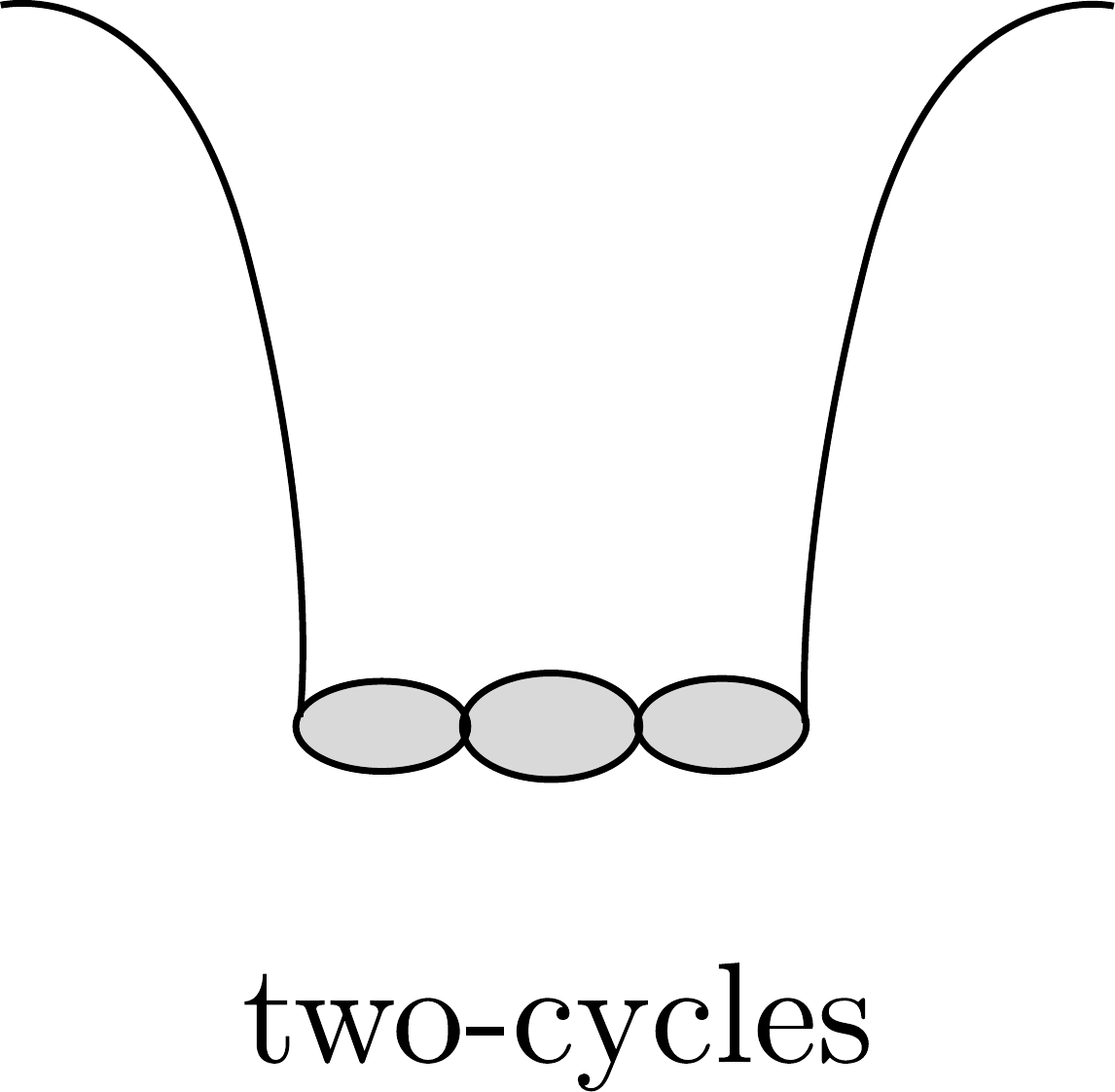}%
 \caption{Top left: the conifold, top right: the deformed conifold. Bottom left: black hole, bottom right: microstate geometry.}
 \label{fig:throat}
\end{figure}

Locally, warped throats of Calabi-Yau spaces are often modeled by the conifold geometry. The conifold describes an infinite cone over a base that is locally $S^2 \times S^3$, and it ends in a conical singularity. However, the singularity at the bottom of the conifold throat does not describe the correct IR phase of the dual field theory. Klebanov and Strassler showed how a duality cascade reveals that the supergravity description of the IR phase  is a singularity-free and smooth solution. The singularity is  resolved by a transition to non-trivial  topology: at the bottom, the $S^3$ (`A-cycle') has a finite size \cite{Klebanov:2000hb}. 
See Figure \ref{fig:throat}, top line. 
%

This type of singularity resolution is very common in string theory, and we know of many (mainly supersymmetric) examples. 
This mechanism is often dubbed ``charge dissolved in flux'', from Chern-Simons like couplings as for instance in IIB, $d \star F_5 = F_3 \wedge H_3$, that allow D3 charge supported by topological fluxes.

Recent ideas hint that the correct IR description of black holes is not an effective field theory near the horizon \cite{Braunstein:2009my,Mathur:2009hf,Almheiri:2012rt}. The microstate geometry programme gives a very concrete suggestion how this could go \cite{Mathur:2008nj,Bena:2013dka}.  Smooth, horizonless microstate geometries have the same leading asymptotics as the  black hole, but non-trivial topology replaces the would-be black hole singularity. Fluxes supported by topological cycles carry the black hole mass and charges and support the system from gravitational collapse, regardless of supersymmetry \cite{Gibbons:2013tqa,Haas:2014spa,deLange:2015gca}. For instance in five dimensions, microstate geometries are supported by fluxes on two-cycles  (Fig.\ \ref{fig:throat}, bottom line).

\subsection{non-SUSY: Smarr relation}\label{ssec:Smarr}

The mechanism to hold up matter with topology and flux is not restricted to supersymmetry, as a careful consideration of the Smarr relation  reveals. 

For the purpose of this note we will continue with black holes in five non-compact spacetime dimensions. The mass is expressed as a Komar integral over the boundary at infinity  ($\partial \Sigma_{\infty} = S^3$): $M =\int_{\partial \Sigma_\infty} \star dK$, with $K$ a timelike Killing vector.  It can be re-expressed as an integral over an interior boundary $\partial \Sigma_{\rm int}$ (typically horizon contribution) but there is an additional bulk term over the interior $\Sigma$ \cite{Gibbons:2013tqa}:
\begin{equation}
M_ = \int_{\partial \Sigma_{\rm int}} {\cal B} + \int_{\Sigma} H \wedge F\,,
\end{equation}
where $F$ is the two-form gauge field and ${\cal B}$ a boundary term containing the fields of the theory. The harmonic form  $H$ is roughly the harmonic part of the dual gauge field contracted with the Killing vector $H = i_K\star  F$.

The Smarr relation shows two possible ways to support mass, irrespective of whether there is supersymmetry. In absence of horizons, only topological terms $\int_\Sigma  H \wedge F$ can support smooth horizonless  microstate geometries. Hence contrary to earlier beliefs, stationary solitons do not need a horizon to support a non-zero mass.

For solutions with  trivial topology, only the boundary term contributes. Schematically it consists of ${\cal B} = \star d K + (i_K A )\star F + i_K F \wedge F$. These terms give
\begin{equation}
 M = T S  + \Omega J + \Phi Q + \phi q\,,\label{eq:MADM_5d}
\end{equation}
with temperature $T$ and entropy $S$, angular velocity $\Omega$ and angular momentum $J$ and the electric charge $Q = \int_{S^3} \star F$ with electrostatic potential $\Phi$. The last term is a contribution of magnetic dipole charge $q = \int_{S^2} F$, with its `dipole potential' $\phi$. The dipole charge $q$ is not a conserved charge and it might seem uprising to see it enter the  mass formula. For instance for dipole black rings one can only define the dipole  charges locally, by integrating the flux over an $S^2$ that links the black ring horizon. The term $\phi q$ appears  due to the inability to define the gauge potential $A$ in a single patch. The dipole contribution comes from the overlap of different patches \cite{Copsey:2005se}. We explain below how a similar effect appears for NS5 dipoles.

\section{Probe branes in flux backgrounds}\label{sec:probes}

Large families of stationary  solutions that break the supersymmetry of throats supported by flux are found by placing probe branes in supersymmetric flux backgrounds. Under the influence of the background flux these probe branes polarize: branes of higher dimensions are generated by blowing up into a contractible cycle. The higher-dimensional brane charge is not a global charge, but rather an induced dipole.

From the point of view of the original probe branes, the polarization stems from a non-commutative effect of the world-volume fields. One can also view this mechanism from the point of view of the polarized brane itself: then the lower-dimensional original brane charge is induced by world-volume flux.

The probe potential of the polarized state allows meta\-sta\-ble minima. These are the supersymmetry-breaking configurations we are after. We review them for flux throats and black hole microstate geometries.

\subsection{Anti-brane probes in flux throats}

\begin{figure}[ht!]
\centering
 \includegraphics[width=.47\columnwidth]{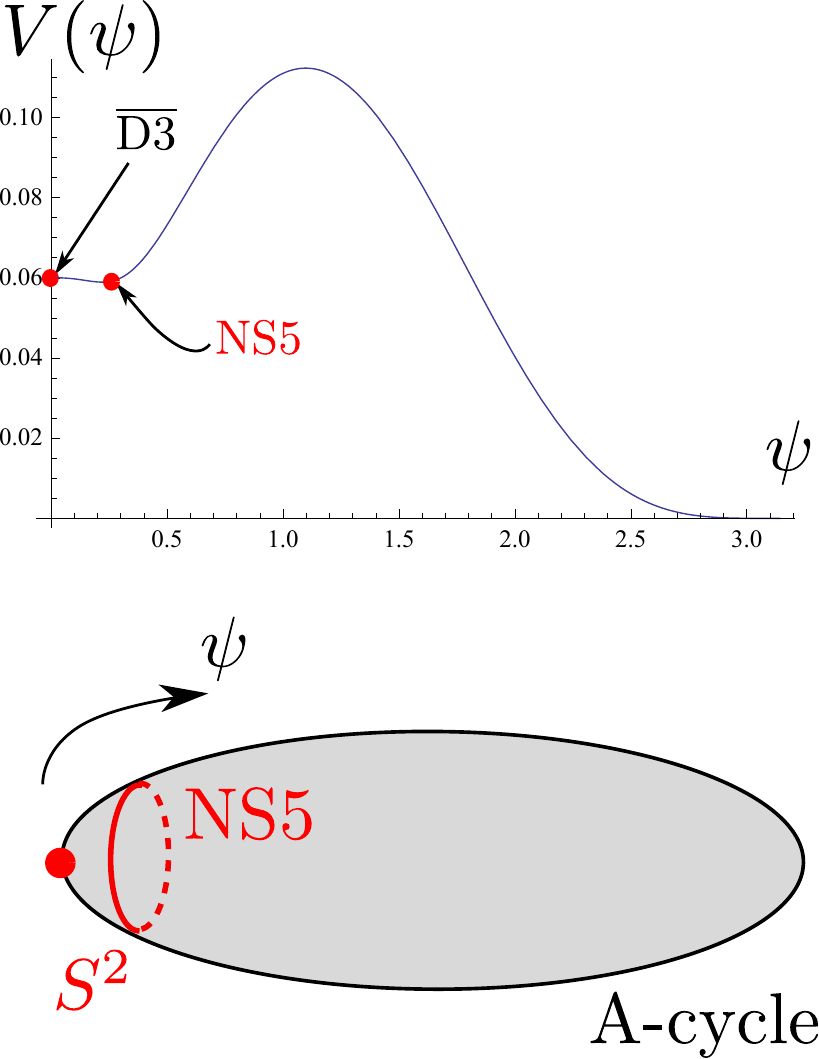}
 \hspace{.02\columnwidth}
  \includegraphics[width=.49\columnwidth]{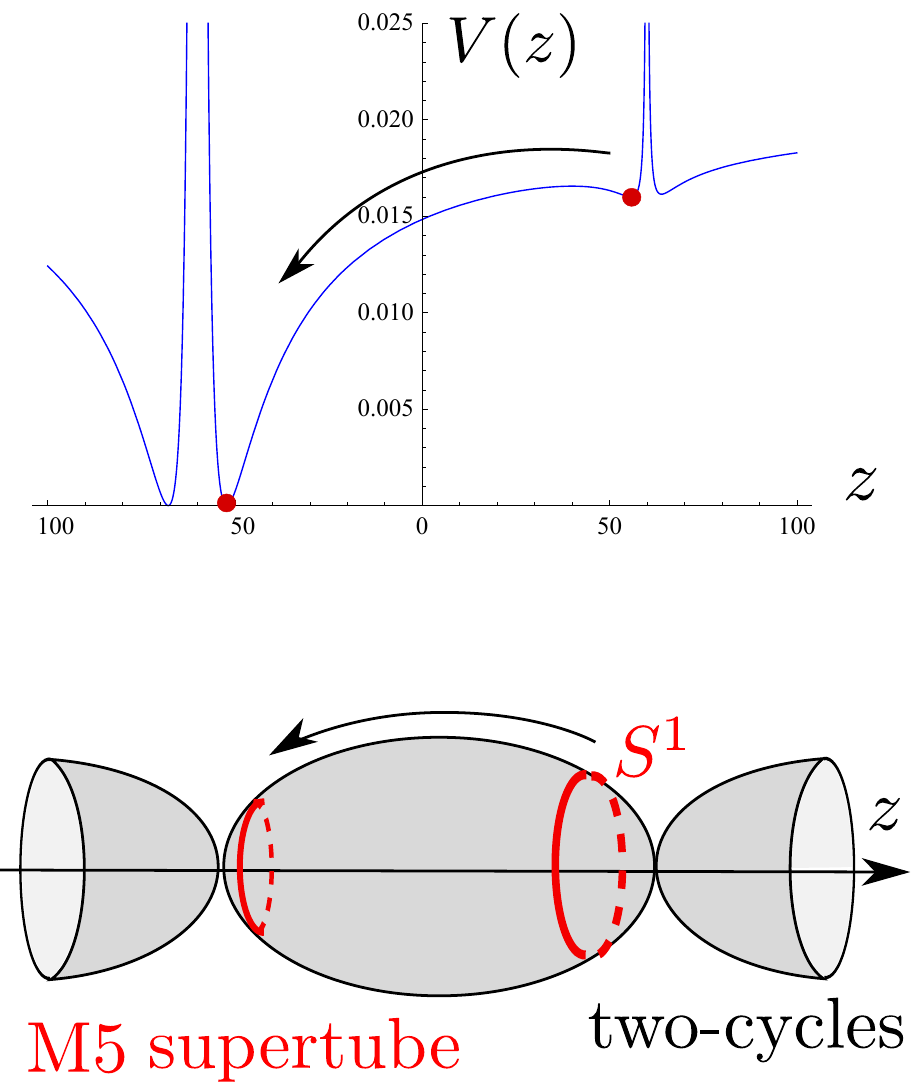}
  \caption{Probe potentials. Left: anti-D3 branes in the Klebanov Strassler background polarizing into NS5 branes. Right: Supertubes as polarized branes in  a supersymmetric microstate geometry. (Note: cycles adapted for illustrative purposes)}
  \label{fig:probe_comparison}
\end{figure}

We zoom in on the warped throat region of the deformed conifold. At the bottom of the throat we find the non-shrinking $S^3$ of the A-cycle. The geometry supports fluxes 
\begin{equation}
\frac 1 {4\pi^2}\int_A F_3 =M\,,\qquad \frac 1 {4\pi^2}\int_B H_3 = K\,,
\end{equation}
with the B-cycle homologically dual to the A-cycle.  The non-zero flux $H_7 \equiv e^{-\phi} \star_{10} H_3$ induces a Myers effect that polarizes $p$ probe anti-D3 branes into dipole NS5-branes that carry anti-D3 brane charge $p$. The NS5-branes extend along the four-external dimensions (the world-volume of the anti-D3 branes) and an extra $S^2$ in the A-cycle. This  has been studied in great detail \cite{Kachru:2002gs}. For $p/M < 0.08$, one finds metastable minima, see figure \ref{fig:probe_comparison}, left. These states were a stepping stone for the de Sitter vacua of  KKLT \cite{Kachru:2003aw}. 

\subsection{Anti-brane probes in microstate backgrounds}

As microstate geometries do not carry entropy (they are horizonless), one would like to add supersymmetry-breaking branes that again do not carry any entropy--such that the total solutions can be interpreted as entropy-less microstates. Therefore we add supertube probes, entropy-less $1/4$  BPS bound states of two types of brane charge that also carry higher-dimensional dipole charge. 

In the M-theory frame the three-charge black hole carries three electric charges from M2 branes wrapped on different two-cycles of the compactification manifold $T^6$. The supertube probes are two types of M2-branes wrapping different two-cycles in $T^6$ that polarize into M5-branes. In five dimensions, the M2-branes are point-like and the M5-branes extend in one spatial dimension (Fig.\ \ref{fig:probe_comparison}).
%
%

Supertubes can sit at supersymmetric and non-super\-sym\-metric positions \cite{Bena:2011fc,Bena:2012zi}. Note that there is one big difference with the anti-branes in KS: due to electromagnetic repulsion between the supertubes and the background centers, the supertubes cannot settle on top of one of the background centers (the equivalent  of the north- and south poles of the $S^3$ in the KS throat). So the metastable state and the supersymmetric end state are \emph{both} dipoles.

\section{Backreaction and polarization}\label{sec:polarize}

The nature of the backreaction of probe branes has been a lively topic over the past 6 years, starting with \cite{McGuirk:2009xx,Bena:2009xk}. For a large enough number of anti-branes $g_s p\gg 1$, such that supergravity is valid, many no-go results have been put forward that point at a singular back-reaction. More recently, very convincing arguments have been put on the table that at least for a single anti-brane and $g_s p < 1$, anti-branes in flux backgrounds would backreact to well-behaved string theory solutions \cite{Michel:2014lva,Bergshoeff:2015jxa,Polchinski:2015bea}.

\subsection{Importance of polarization}
We believe that in this debate, the fate of the polarized state has not been given enough attention. The polarized brane, and not the original anti-brane, is metastable in the probe analysis--both for NS5-branes in KS and for supertubes in microstate backgrounds. However, a lot of the literature has focused either on unpolarized anti-branes, or on polarized states in limits where the original probe analysis of \cite{Kachru:2002gs} did not observe polarization at all, such as branes smeared over the A-cycle or the regime $p/M \gg 1$. 

In this note and in \cite{Cohen-Maldonado:2015ssa}, we study the polarized state in the supergravity regime. We choose to focus on $g_s p \gg 1$ partly because of control of calculation, and also because of the extreme importance in light of a possible geometric nature of black hole microstates. 

\subsection{Applying the Smarr relation}

The Smarr relation of section \ref{ssec:Smarr} is an enormously powerful tool to study anti-brane back-reaction. It has been derived in \cite{Gautason:2013zw} to make statements about the back-reaction of unpolarized D3-states in compact throats, and applied to non-compact throats with $T>0$ in \cite{Blaback:2014tfa}. The power of this approach is that one does not need the details of the back-reaction. By the Smarr relation one can match the expected asymptotic behaviour (the ADM mass for anti-D3 branes in KS) to the known near-brane behaviour of anti-branes. This has lead to the conclusion that the back-reaction would have an unphysical singularity in the $H$-flux density:  $e^{-\phi} |H_3|^2 \to \infty$ at the position of the anti-branes.

However, the polarized state has not been studied with the method of \cite{Gautason:2013zw,Blaback:2014tfa}. The relevant part of the boundary term $\cal B$ entering the Smarr relation is
\begin{equation}
M = \int_{\partial {\cal M}_{\rm int}} {\cal B}\,,\qquad  {\cal B} = - C_4 \wedge F_5 - B_6 \wedge H_3\,,
\end{equation}
where the integration is now over the inner boundary of ten-dimensional piece of spacetime $\cal M$ that encapsulates the anti-branes. Letting $\partial {\cal M}_{\rm int}$ be infinitesimally close to the NS5-brane, one finds (compare eq.\ \ref{eq:MADM_5d})
\begin{equation}
 M = \alpha_H Q_3 + b_H Q_5 V_{S^2} >0\,,\label{eq:MADM_KS}
\end{equation}
with the anti-D3 monopole charge and NS5 dipole charge defined as integrals over spheres in the 6d internal space: $Q_3 = \int_{S^5} F_5$, $Q_5 = \int_{S^3} H_3$, and $\alpha_H$ and $b_H$ the constant values of the D3 and NS5 gauge fields at the position of the anti-branes; $V_{S^2}$ is the volume of the polarization two-sphere. We used that the mass is proportional to the number of anti-branes: $M\propto p>0$.

We can confront the non-zero mass with the near-brane behaviour of the flux density. If we require that $e^{-\phi} |H_3|^2$ blows up in the expected way to be the NS5-brane self-energy, one finds that near the anti-branes
\begin{equation}
 \alpha_H \to 0, \qquad \partial_r b_H \to 0\,,\label{eq:flux_density_constraint}
\end{equation}
where $r$ is a radial coordinate transverse to the NS5-brane.  Comparing \eqref{eq:MADM_KS} and \eqref{eq:flux_density_constraint} does not give an obstruction to well-behaved NS5 back-reaction.
In absence of an NS5 polarization channel however, $b_H=0$ and  \eqref{eq:MADM_KS} and \eqref{eq:flux_density_constraint} cannot be satisfied simultaneously, explaining previous no-go results. 

We do want to point out that, as explained in our work \cite{Cohen-Maldonado:2015ssa}, the Smarr relation indicates that the volume of the $S^2$ at which the NS5 settles is linear in the anti-brane charge $V_{S^2} \propto p/M$, while the probe treatment suggests it is quadratic $V_{S^2} \propto (p/M)^2$. Since we take $p/M\ll 1$, the back-reaction pushes the anti-branes further away: a clear sign of the possibly dangerous effect of flux clumping  \cite{Blaback:2012nf}. This difference in dependence on the probe charge suggests an issue with the consistency of the probe limit. It would be very interesting to match our results to an EFT treatment of the NS5 polarization along the lines of \cite{Polchinski:2015bea}.  

We plan to extend the analysis of the Smarr relation to more general KS asymptotics that allow $p/M \gg 1$ and to the context of non-supersymmetric black hole microstates. This  will allow to test the recent findings that black hole microstate geometries obtained from  probe anti-branes should be unstable \cite{Bena:2015lkx}.

\section*{Acknowledgments}

We acknowledge support from the European Science Foundation Holograv Network. TVR is supported by the National Science Foundation of Belgium (FWO) grant G.0.E52.14N Odysseus. The research of BV is supported by the European Commission through the Marie Curie Intra-European fellowship 328652--QM--sing. DCM would like to acknowledge the Becas Chile scholarship programme of the Chilean government.

\bibliographystyle{toine}
\bibliography{prop_NS5.bib}

%
%
%
%
%
%
%
%
%
\end{document}